\documentclass[aps,twocolumn,showpacs,floatfix,amssymb]{revtex4}
\usepackage{graphicx}
\begin{document}

\title{Magnetic order in the two-dimensional compass-Heisenberg model}

\author{ A.A. Vladimirov$^{a}$, D. Ihle$^{b}$ and  N. M. Plakida$^{a}$ }
 \affiliation{ $^a$Joint Institute for Nuclear Research,
141980 Dubna, Russia}
 \affiliation{$^{b}$ Institut f\"ur Theoretische Physik,
 Universit\"at Leipzig,  D-04109, Leipzig, Germany }

\date{\today}

\begin{abstract}
A Green-function theory for the dynamic spin susceptibility in
the square-lattice spin-$1/2$ antiferromagnetic
compass-Heisenberg model employing a generalized mean-field
approximation is presented. The theory describes magnetic
long-range order (LRO) and short-range order (SRO) at arbitrary
temperatures. The magnetization, N\'{e}el temperature $T_N$,
specific heat, and uniform static spin susceptibility $\chi$ are
calculated self-consistently. As the main result, we obtain LRO
at finite temperatures in two dimensions, where the dependence of
$T_N$ on the compass-model interaction is studied. We find that
$T_N$ is close to the experimental value for Ba$_2$IrO$_4$. The
effects of SRO are discussed in relation to the temperature
dependence of $\chi$.

{Published in: Eur. Phys.  J. {\bf B 88}, 148 (2015).}
\end{abstract}


\maketitle

\section{Introduction}
The study of the quantum compass model (CM) for strongly
correlated transition-metal compounds with orbital degrees of
freedom and strong spin-orbit coupling is an active field of
research (for a review, see Ref.~\cite{NB15}). In particular,
quantum and thermodynamic phase transitions in the
two-dimensional (2D) CM were studied in
Refs.~\cite{Wenzel08,Wenzel0}, where the directional-ordering
transition of the 2D Ising universality class was found for the
symmetric CM. Depending on the method of numerical computation,
the temperature of the phase transition for the quantum CM was
found in the range $T_c = 0.055 - 0.058$ (in the units of the CM
exchange interaction)~\cite{Wenzel0}. In Refs.~\cite{TOH10,TOH12}
the compass-Heisenberg (CH) model was introduced extending the CM
by the 2D antiferromagnetic (AF) Heisenberg model, and the ground
state and excited states of the CH model were analyzed. For 5d
transition-metal compounds with a strong spin-orbit coupling,
such as Sr$_2$IrO$_4$ and Ba$_2$IrO$_4$ with the N\'{e}el
temperatures $T_N = 230$K and $T_N = 240$K, respectively, (see,
e.g., Ref.~\cite{BSW13}), an effective AF Heisenberg model for
pseudospins $S = 1/2$ with the CM anisotropy was derived in
Ref.~\cite{JK09}. Recently we have calculated the spin-wave
excitation spectrum and $T_N$ for a layered CH model by means of
the random phase approximation (RPA)~\cite{VIP14} (see also
Refs.~\cite{Igarashi13}). An important issue is the description
of magnetic long-range order (LRO) and short-range order (SRO)
and of the thermodynamics at arbitrary temperatures by a theory
going beyond RPA.

In this paper we employ a generalized mean-field approximation
(GMFA) to the 2D CH model that is based on the equation-of-motion
method for Green functions. In the framework of a more general
theory for the dynamic spin susceptibility including the self
energy~\cite{VIP05,VIP09}, the neglect of the self-energy
corresponds to the GMFA. This approximation, for spin-rotation
invariant (SRI) systems also named SRI Green-function method
(RGM), has been successfully applied to several quantum spin
systems (see, e.g., Refs.
\cite{KY72,ST91,BB94,WI97,ISW99,YF00,BCL02,SDR06,HRI08,JIB08,JIR09,MKB09,BMS11,HRG13,VIP14a}).

We start from the spin-$1/2$ CH model on the square lattice,
\begin{equation}
H = \frac{1}{2} \sum_{i,j,\nu} J^{\nu}_{ij} S^{\nu}_i S^{\nu}_j,
\label{eq1}
\end{equation}
where $\nu = x, y, z$. The nearest-neighbor (NN) exchange interaction
parameters are $J^x_{ij} = J_{ij} + \Gamma^x_{ij}$,
$J^y_{ij} = J_{ij} + \Gamma^y_{ij}$, $J^z_{ij} = J_{ij}$,
$J_{ij} = J (\delta_{{\bf r}_j, {\bf r}_i \pm {\bf a}_x} +
\delta_{{\bf r}_j, {\bf r}_i \pm {\bf a}_y})$,
$\Gamma^x_{ij} = \Gamma_x \delta_{{\bf r}_j, {\bf r}_i \pm {\bf a}_x}$, and
$\Gamma^y_{ij} = \Gamma_y \delta_{{\bf r}_j, {\bf r}_i \pm {\bf a}_y}$.
We assume $J > 0$ and $\Gamma_x \geq \Gamma_y > 0$.
The symmetric formulation of the model (\ref{eq1}) allows us to get
expressions for quantities with indexes $\nu = y,$ $z$ from those
indicated by the index $\nu  = x$ by cyclic permutation.

\section{Theory of spin susceptibility}

To evaluate the thermodynamic quantities in the CH model, we calculate the
dynamic spin susceptibility $\chi^{\nu}_{\bf q} (\omega) =
 -\langle\langle S^{\nu}_{\bf q} | S^{\nu}_{\bf q} \rangle\rangle_{\omega}$
($\langle\langle ...|... \rangle\rangle_{\omega}$ denotes the retarded
two-time commutator Green function~\cite{T67}). Using the equations of
motion up to the second step, we obtain
$\omega^2 \langle\langle S^{\nu}_{\bf q} | S^{\nu}_{\bf q}
\rangle\rangle_{\omega} = m^{\nu}_{\bf q} + \langle\langle
-\ddot{S}^{\nu}_{\bf q} | S^{\nu}_{\bf q} \rangle\rangle_{\omega}$,
where $m^{\nu}_{\bf q} = \langle [i \dot{S}^{\nu}_{\bf q} |
S^{\nu}_{\bf q}] \rangle$,
$i\dot{S}^{\nu}_{\bf q} = [S^{\nu}_{\bf q}, H]$, and
 $-\ddot{S}^{\nu}_{\bf q} = [[S^{\nu}_{\bf q}, H], H]$.
For the model (\ref{eq1}) the
moment $m^x_{\bf q}$ is given by the exact expression
\begin{equation}
m^x_{\bf q} = \sum_{i} [\cos ({\bf qR}_i)(J^y_{0i}C^z_{0i} + J^z_{0i}C^y_{0i})
- J^y_{0i} C^y_{0i} - J^z_{0i} C^z_{0i}].
\label{eq2}
\end{equation}
Here, $C^{\nu}_{ij} = C^{\nu}_{{\bf R}_j - {\bf R}_i} =
\langle S^{\nu}_0 S^{\nu}_{{\bf R}_j - {\bf R}_i} \rangle$ with ${\bf R} =
 m{\bf e_x} + n{\bf e_y}$ denote the spin correlation functions.

The second derivatives $-\ddot{S}^{\nu}_{\bf q}$ are approximated
in the spirit of the scheme employed in
Refs.~\cite{KY72,ST91,BB94,WI97,ISW99,YF00,BCL02,SDR06,HRI08,JIB08,JIR09,MKB09,BMS11,HRG13,VIP14a}.
That means, taking the site representation, in $-\ddot{S}^x_i$ we
decouple the products of three spin operators on different
lattice sites along NN sequences $\langle i, j, k \rangle$ as
\begin{eqnarray}
S^x_i S^y_j S^y_k = \alpha^x_1 \langle S^y_j S^y_k \rangle S^x_i,\label{eq3}\\
S^x_j S^y_i S^y_k = \alpha^x_2 \langle S^y_i S^y_k \rangle S^x_j,\label{eq4}
\end{eqnarray}
where the vertex renormalization parameters $\alpha^x_1$ and $\alpha^x_2$ are
attached to NN and further-distant correlation functions, respectively.
After some algebra, we obtain $-\ddot{S}^{\nu}_{\bf q} =
(\omega^{\nu}_{\bf q})^2 S^{\nu}_{\bf q}$
and
\begin{equation}
\chi^{\nu}_{\bf q} (\omega) =
-\langle\langle S^{\nu}_{\bf q} | S^{\nu}_{\bf q} \rangle\rangle_{\omega}=
\frac{m^{\nu}_{\bf q}}{(\omega^{\nu}_{\bf q})^2 - \omega^2},
\label{eq5}
\end{equation}
with the squared spin-excitation energy
\begin{equation}
(\omega^x_{\bf q})^2 = \sum_{i,j} [\delta_{ij} a^x_i({\bf q}) +
(1-\delta_{ij}) b^x_{ij}({\bf q})], \label{eq6}
\end{equation}
where
\begin{equation}
a^x_i({\bf q})=
\frac{1}{4}[
(J^y_{0i})^2 + (J^z_{0i})^2
- 2J^y_{0i} J^z_{0i} \cos ({\bf qR}_i)],
\label{eq7}
\end{equation}
\begin{eqnarray}
b^x_{ij}({\bf q})&=&
  \alpha^x_2 J^y_{0i} J^y_{0j} C^y_{ij}
+ \alpha^x_2 J^z_{0i} J^z_{0j} C^z_{ij}\nonumber\\
&&+\alpha^x_1 \cos ({\bf qR}_{ij})
(
 J^y_{0i} J^x_{0j} C^z_{0i}
+J^z_{0i} J^x_{0j} C^y_{0i}
)\nonumber\\
&&-\cos ({\bf qR}_i)
(
 \alpha^x_1 J^x_{0i} J^y_{0j}  C^y_{0j}
+\alpha^x_1 J^x_{0i} J^z_{0j}  C^z_{0j}\nonumber\\
&&+\alpha^x_2 J^y_{0i} J^z_{0j}  C^z_{ij}
+\alpha^x_2 J^z_{0i} J^y_{0j}  C^y_{ij}
).
\label{eq8}
\end{eqnarray}

The appearance of AF LRO at $T \leq T_N$ is reflected in our theory
by the divergence of $\chi^{\nu}_{\bf Q} \equiv \chi^{\nu}_{\bf Q}(\omega = 0)$
corresponding to the closure of the spectrum gap at the AF ordering vector
${\bf Q} = (\pi, \pi)$, $\omega^{\nu}_{\bf Q} (T \leq T_N) = 0$. In the
LRO phase the correlation functions $C^{\nu}_{\bf R} \equiv C^{\nu}_{m,n}$
are written as~\cite{ST91,WI97,ISW99,JIR09,HRG13,VIP14a}
\begin{equation}
C^{\nu}_{\bf R} = \frac{1}{N}\sum_{{\bf q} \neq {\bf Q}}
C^{\nu}_{\bf q} e^{i \bf qR} + C^{\nu}e^{i \bf QR},
\label{eq10}
\end{equation}
with $C^{\nu}_{\bf q}$ calculated from the Green function (\ref{eq5}) by the
spectral theorem,
\begin{equation}
C^{\nu}_{\bf q} = \langle S^{\nu}_{\bf q} S^{\nu}_{\bf q} \rangle =
\frac{m^{\nu}_{\bf q}}{2\omega^{\nu}_{\bf q}}[1 + 2n(\omega^{\nu}_{\bf q})],
\label{eq11}
\end{equation}
where $n(\omega) = (e^{\omega / T} - 1)^{-1}$.
The condensation part $C^{\nu}$ arising from
$\omega^{\nu}_{\bf Q} = 0$ determines the staggered magnetization
$m^{\nu}$ that is defined by
\begin{equation}
(m^{\nu})^2 = \frac{1}{N}\sum_{\bf R}C^{\nu}_{\bf R} e^{-i \bf QR} = C^{\nu}.
\label{eq12}
\end{equation}
In the paramagnetic phase, we have $\omega^{\nu}_{\bf Q} > 0$ and
$C^{\nu} = 0$. The NN correlation functions are related to the internal
energy $u$ per site,
$u = (1 / 2N) \sum_{i,j,\nu} J^{\nu}_{ij} C^{\nu}_{ij}$,
from which the specific heat $C_V = du/dT$ may be obtained.

To calculate the thermodynamic properties, the correlation
functions $C^{\nu}_{\bf R}$, the vertex parameters
$\alpha^{\nu}_{1,2}$ appearing in the spectrum $\omega^{\nu}_{\bf
q}$, and the condensation term $C^{\nu}$ in the LRO phase have to
be determined. Besides Eqs. (\ref{eq10}) and (\ref{eq11}) for
calculating the correlators, we have the sum rules $C^{\nu}_{{\bf
R} = 0} = 1/4$ and the LRO conditions $\omega^{\nu}_{\bf Q} = 0$;
that is, we have more parameters than equations. To obtain a
closed system of self-consistency equations, we proceed as
follows. As in our recent study of the model (\ref{eq1}) by means
of the RPA and linear spin-wave theory (LSWT)~\cite{VIP14}, we
consider an anisotropic CM interaction, $\Gamma_x > \Gamma_y > 0$,
so that the LRO phase is an easy-axis AF with the magnetization
along the x axis. Accordingly, we put $C^y = C^z = 0$, where we
can also consider the limiting case $\Gamma_x = \Gamma_y$. Then ,
at T = 0, for determining the seven quantities
$\alpha^{\nu}_{1,2}$ and $C^x$, besides the three sum rules and
the three LRO conditions, we need an additional condition. To
this end, we adjust the ground-state magnetization $m^x(0)$ to the
expression obtained in LSWT,
 $m^x(0) = m^x_{LSWT}(0) = (1 / 2) - (1 / N)\sum_{\bf q}
 \langle S^-_{\bf q} S^+_{\bf q} \rangle$,
where the correlation function is given
by Eq. (17) of Ref.~\cite{VIP14} with the sublattice magnetization
$\sigma$ substituted by the spin $S = 1/2$.
In the LRO phase, $0 < T \leq T_N$, we have found that the
ansatz $\bar{r}_{\alpha}(T) = \bar{r}_{\alpha}(0)$, where
$\bar{r}_{\alpha} = (r^x_{\alpha} r^y_{\alpha} r^z_{\alpha})^{1/3}$ with
$r^{\nu}_{\alpha} (T) = (\alpha^{\nu}_2 (T) - 1) / (\alpha^{\nu}_1 (T) - 1)$,
is a reasonable approximation, as will be demonstrated in the Appendix
for the Heisenberg limit. At $T > T_N$, for calculating
$\alpha^{\nu}_{1,2}$ we use the sum rules and the
condition $r^{\nu}_{\alpha} (T) = r^{\nu}_{\alpha} (T_N)$.

Because, as an input, we take the ground-state magnetization in
LSWT that describes the LRO quite well for small enough CM
interaction $\Gamma_{x, y}$ as compared to the exchange
interaction $J$, we present   results  in the region $\Gamma_{x,
y}/J \ll 1$.

\section{Results}

As described in Sec. II, the thermodynamic quantities are calculated from
the numerical solution of a coupled system of nonlinear algebraic
self-consistency equations.

\begin{figure}
\resizebox{0.38\textwidth}{!}{%
\includegraphics{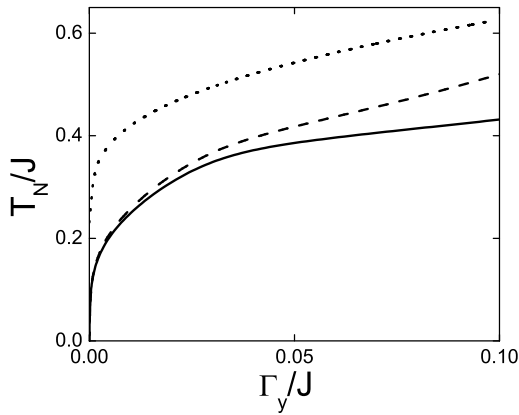}}
\caption{N\'{e}el temperature $T_N$ for $\Gamma_x = \Gamma_y$
(solid) and $\Gamma_x = 1.5 \Gamma_y$ (dashed) compared with the
RPA result (dotted).}
 \label{fig1}
\end{figure}

First we consider our results on the magnetic LRO. In the
Heisenberg limit we have no LRO at finite temperature, in accord
with the Mermin-Wagner theorem. In Fig. \ref{fig1} the N\'{e}el
temperature $T_N$ as function of $\Gamma_y$ for two ratios of
$\Gamma_x / \Gamma_y$ is plotted. Our results for $T_N$
remarkably deviate from those found in RPA~\cite{VIP14}. For
$\Gamma_x / \Gamma_y = 1.5$, $T_N$ exhibits qualitatively the
same dependence on $\Gamma_y$, but is appreciably reduced as
compared to RPA. For the symmetric CM interaction $\Gamma_x =
\Gamma_y = \Gamma$, our theory yields a finite N\'{e}el
temperature that increases with $\Gamma$, whereas in RPA, $T_N =
0$ was found. We ascribe these differences to a better
description of strong spin fluctuations by our Green-function
theory going one step beyond RPA. Note that in the
RPA~\cite{VIP14} the fluctuations of the $x$ component of spin
are not taken into account. Generally speaking, from our results
we conclude that the CM interaction added to the Heisenberg model
favors magnetic LRO at finite $T$ in two dimensions.

\begin{figure}
\resizebox{0.38\textwidth}{!}{%
\includegraphics{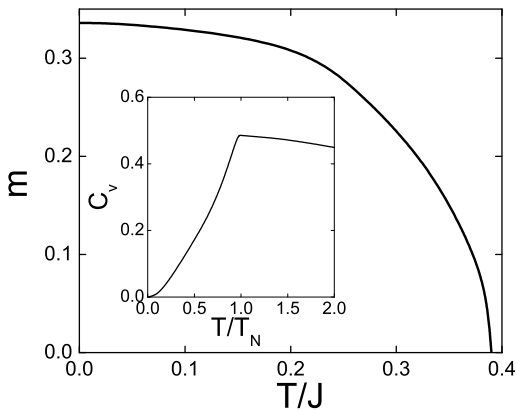}}
\caption{Magnetization $m$ and specific heat $C_V$ for $\Gamma_x
= \Gamma_y = 0.0523 J$.}
 \label{fig2}
\end{figure}

Comparing our findings to experiment, we consider the compound Ba$_2$IrO$_4$,
where quantum chemistry calculations~\cite{KYS14} yield the parameters
$J=65$meV, $\Gamma = 3.4$meV $= 0.0523 J$, and a very small interplane
coupling $J_z \approx 5\times 10^{-5} J$ that is not taken into account here.
In Fig. \ref{fig2} the magnetization $m = m^x$ as function of $T$ is plotted.
At $T=0$, $m$ is enhanced by $\Gamma$ as compared to the value in the
Heisenberg limit ($m_{LSWT}(0) = 0.303$). The second-order phase transition
occurs at $T_N = 0.392 J = 295$K which agrees rather well with the
experimental value $T_N^{exp} = 240$K. At $T_N$ the specific heat
(inset to Fig. \ref{fig2}) reveals a cusplike singularity.
Such a structure at the transition temperature
was also found in layered magnets treated by RGM in Ref.~\cite{JIR09}.
From the analysis given there we suggest that also here the height of the
cusp may be underestimated. It is desirable to compare our result for $C_V$
to experimental data which, however, are not yet available.

\begin{figure}
\resizebox{0.38\textwidth}{!}{%
\includegraphics{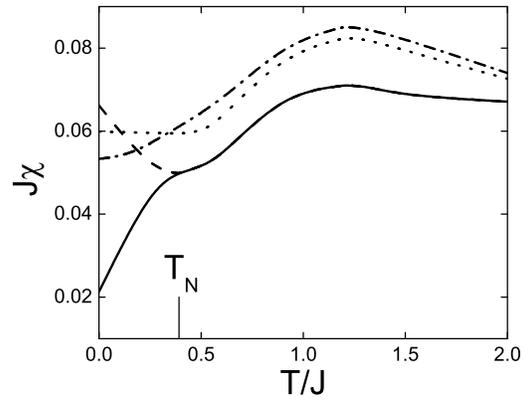}}
\caption{Uniform static susceptibility $\chi$ for $\Gamma_x =
\Gamma_y = 0.0523 J$, $\chi^x$ (solid), $\chi^y$ (dashed),
$\chi^z$ (dotted), and for the Heisenberg limit (dot-dashed).}
 \label{fig3}
\end{figure}

Finally, we discuss the uniform static spin susceptibility
$\chi(T) \equiv \chi_{{\bf q} = 0}(\omega = 0)$ depicted in Fig.
\ref{fig3} for the symmetric CM interaction, which reflects the
behavior of magnetic SRO. Considering the Heisenberg limit, the
increase of $\chi$ with temperature is caused by the decrease of
AF SRO, i.e., of the spin stiffness against the orientation along
a homogeneous magnetic field. Since the SRO is less pronounced at
higher $T$, $\chi$ reveals a maximum near the exchange energy and
a crossover to the high-temperature Curie-Weiss behavior. In the
paramagnetic phase for $\Gamma > 0$, the susceptibilities
$\chi^{\nu}$ are lowered as compared with the Heisenberg limit
due to the $\Gamma$-induced enhancement of AF SRO. Moreover, the
susceptibility $\chi^x = \chi^y$ is reduced as compared with
$\chi^z$, because the SRO of the x- and y-spin components is more
pronounced due to the AF CM interaction along the x- and y-axes.
In the LRO phase, the susceptibility $\chi^x$ strongly decreases
with decreasing $T$, because the increasing easy-x-axis LRO also
enhances the SRO. Considering the susceptibility $\chi^y$, with
decreasing $T$ below $T_N$ the SRO of the y-spin components is
reduced due to the increasing alignment of the spins in
x-direction, so that $\chi^y$ increases with decreasing $T$.
Thus, $\chi^y$ exhibits a minimum at $T_N$. The susceptibility
$\chi^z$ in the LRO phase turns out to be nearly temperature
independent.

\section{Summary}
In this paper we have calculated thermodynamic properties of the
2D antiferromagnetic CH model within a generalized mean-field
approach for arbitrary temperatures. Our main focus was the
detailed investigation of magnetic LRO. We have found that the
N\'{e}el temperature $T_N$ is finite even in the case of a
symmetric CM interaction. We have explained the temperature
dependence of the uniform static susceptibility in terms of
magnetic SRO. Our investigation forms the basis for forthcoming
extended studies of the CH model.  Especially, the generalization
of the theory holding for arbitrary CH model parameters and
allowing for the description of magnetic and directional-ordering
transitions is of particular interest.\\

The authors would like to thank W. Janke and J. Richter for
valuable discussions. The financial support by the
Heisenberg-Landau program of JINR is acknowledged.

\appendix
\section{Heisenberg limit}

\begin{table}
\caption{Heisenberg limit at $T = 0$. }
 \label{table1}
\begin{tabular}{crrrc}
\hline \hline\\
 & \hspace{0.2cm} $\nu = x$ & \hspace{0.2cm} $\nu = y, z$
& \hspace{0.2cm} average  & \hspace{0.2cm} SRI\\
\hline \hline
\rule{0pt}{3ex} $C_{10}$& -0.1542 & -0.0989 & -0.1173 & -0.1173 \\
\rule{0pt}{3ex} $C_{20}$& 0.118 & 0.0412 & 0.0668 & 0.0667 \\
\rule{0pt}{3ex} $C_{11}$& 0.125 & 0.0522 & 0.0765 & 0.0763 \\
\rule{0pt}{3ex} $J \chi$& 0.0327 & 0.0636 & 0.0533 & 0.05292 \\
\rule{0pt}{3ex} $\Omega / J$& 3.477 & 2.821 & 3.04 & 2.978 \\

\hline \hline
\end{tabular}\\

\end{table}

In the limit $\Gamma_{x,y} = 0$, the LRO in the ground state of
the Heisenberg AF may be described within the RGM by the SRI
forms of Eqs. (\ref{eq10}) and (\ref{eq12}), i.e., by $C^x_{\bf
R} = C^y_{\bf R} = C^z_{\bf R}$ and $C^x = C^y = C^z$ as it was
done, e.g., in Refs.~\cite{ST91,WI97,JIR09,VIP14a}. According to
the consideration of the CH model with the easy-axis magnetization
$m^x$, let us outline an alternative possibility to describe the
ground-state LRO in the AF Heisenberg model. Here, we may also
break the rotational symmetry by putting $C^y = C^z = 0$, which
is analogous to the introduction of a symmetry-breaking
sublattice magnetization (see, e.g., our RPA
approach~\cite{VIP14}). Performing the calculations at $T=0$ as
described in Sec. II, we obtain the correlation functions
$C^{\nu}_{mn}$ and the uniform static spin susceptibility
$\chi^{\nu}$ listed in Table \ref{table1}. The spin-excitation
spectrum at $T=0$ is found to be
\begin{equation}
\omega^{\nu}_{\bf q} = \Omega^{\nu} \sqrt{1-\gamma^2_{\bf q}},
\label{A1}
\end{equation}
where $\gamma_{\bf q} = \frac{1}{2}(\cos q_x + \cos q_y)$ and
$(\Omega^x)^2 = -16J^2\alpha^x_1 (C^y_{10} + C^z_{10})$.
The amplitudes $\Omega^{\nu}$ are given in Table \ref{table1}.
Note that Eq. (\ref{A1}) leads to the SRI result (Ref.~\cite{WI97}),
if we put $\alpha^x_1 = \alpha_1$ and $C^y_{10} + C^z_{10} = 2C^z_{10}$,
and has the same shape as in LSWT and RPA.

To relate the non-SRI description of LRO to the SRI formulation, we calculate
arithmetic averages, e.g., $C_{mn} = \frac{1}{3}\sum_{\nu}C^{\nu}_{mn}$.
As can be seen from Table \ref{table1}, we obtain a very good agreement of
the averaged ground-state properties with those resulting from the SRI theory.
Moreover, the geometrical average $\bar{r}_{\alpha}(0)$ used in
our ansatz for the LRO phase (see Sec. II) given in the Heisenberg limit by
$\bar{r}_{\alpha}(0) = 1.1913$ nearly agrees with the ratio
$r_{\alpha}(0) = 1.2109$ obtained in the SRI approach
(see also Ref.~\cite{ST91}).
This gives some justification for formulating our ansatz in terms of
the geometrical average.


\begin{thebibliography}{99}

\bibitem{NB15} Z. Nussinov and J. van den Brink, Rev. Mod. Phys. {\bf 87},
1 (2015).

\bibitem{Wenzel08}  S. Wenzel and W. Janke, Phys. Rev. {\bf  B  78}, 064402 (2008).

\bibitem{Wenzel0}  S. Wenzel,  W. Janke, and
 A.\,M. L\"{a}uchli,  Phys. Rev. {\bf  E  81}, 066702  (2010).

\bibitem{TOH10} F. Trousselet, A. M. Ole\'{s}, and P. Horsch, Europhys.
Lett. {\bf 91}, 40005 (2010).

\bibitem{TOH12} F. Trousselet, A. M. Ole\'{s}, and P. Horsch, Phys. Rev. B
{\bf 86}, 134412 (2012).

\bibitem{BSW13} S. Boseggia, R. Springell, H. C. Walker, H. M. R{\o}nnow,
Ch. R\"{u}egg, H. Okabe, M. Isobe, R. S. Perry,
S. P. Collins, and D. F. McMorrow, Phys. Rev. Lett.
{\bf 110}, 117207 (2013).

\bibitem{JK09} G. Jackeli and G. Khaliullin, Phys. Rev. Lett. {\bf 102},
017205 (2009).

\bibitem{VIP14} A. A. Vladimirov, D. Ihle, and N. M. Plakida,
JETP Lett. {\bf 100}, 780 (2014), arXiv:1411.3920v2.

\bibitem{Igarashi13} J. I. Igarashi and T.Nagao, Phys. Rev.  {\bf B 88}, 104406
(2013); {\bf B 89}, 064410 (2014).

\bibitem{VIP05}A. A. Vladimirov, D. Ihle, and N. M. Plakida,
Theor. Mat. Phys. {\bf 145}, 1576 (2005).

\bibitem{VIP09}A. A. Vladimirov, D. Ihle, and N. M. Plakida,
Phys. Rev. B {\bf 80}, 104425 (2009); {\bf 83}, 024411 (2011).

\bibitem{KY72} J. Kondo and K. Yamaji, J. Theor. Phys. {\bf 47}, 807 (1972).

\bibitem{ST91} H. Shimahara and S. Takada, J. Phys. Soc. Jpn. {\bf 60},
2394 (1991).

\bibitem{BB94} A. F. Barabanov and V. M. Berezovsky, J. Phys. Soc. Jpn.
{\bf 63}, 3974 (1994);
Phys. Lett. A {\bf 186}, 175 (1994);
Zh. Eksp. Teor. Fiz. {\bf 106}, 1156 (1994) [JETP {\bf 79}, 627 (1994)].

\bibitem{WI97} S. Winterfeldt and D. Ihle, Phys. Rev. B {\bf 56}, 5535 (1997).

\bibitem{ISW99} D. Ihle, C. Schindelin, and H. Fehske, Phys. Rev B {\bf 64},
054419 (2001).

\bibitem{YF00} W. Yu and S. Feng, Eur. Phys. J. B {\bf 13}, 265 (2000).

\bibitem{BCL02} B. H. Bernhard, B. Canals, and C. Lacroix, Phys. Rev. B
{\bf 66}, 104424 (2002).

\bibitem{SDR06} D. Schmalfu\ss, R. Darradi, J. Richter, J. Schulenburg,
and D. Ihle, Phys. Rev. Lett. {\bf 97}, 157201 (2006).

\bibitem{HRI08} M. H\"{a}rtel, J. Richter, D. Ihle, and S.-L. Drechsler,
Phys. Rev. B {\bf 78}, 174412 (2008); {\bf 81}, 174421 (2010).

\bibitem{JIB08} I. Juh\'{a}sz Junger, D. Ihle, L. Bogacz, and W. Janke,
Phys. Rev. B {\bf 77}, 174411 (2008).

\bibitem{JIR09} I. Juh\'{a}sz Junger, D. Ihle, and J. Richter,
Phys. Rev. B {\bf 80}, 064425 (2009).

\bibitem{MKB09} A. V. Mikheenkov, N. A. Kozlov, and A. F. Barabanov,
Phys. Lett. A {\bf 373}, 693 (2009).

\bibitem{BMS11} A. F. Barabanov, A. V. Mikheenkov, and A. V. Shvartsberg,
Theor. Math. Phys. {\bf 168}, 1192 (2011).

\bibitem{HRG13} M. H\"{a}rtel, J. Richter, O. G\"{o}tze, D. Ihle,
and S.-L. Drechsler, Phys. Rev. B {\bf 87}, 054412 (2013).

\bibitem{VIP14a} A. A. Vladimirov, D. Ihle, and N. M. Plakida,
Eur. Phys. J. B {\bf 87} 112 (2014).

\bibitem{T67} S. V. Tyablikov, Methods in the Quantum Theory of
Magnetism, Plenum, N.Y. (1967) [2-nd Edition, Nauka,
M. (1975)].

\bibitem{KYS14} V. M. Katukuri, V. Yushankhai, L. Siurakshina, J. van den
Brink, L. Hozoi, and I. Rousochatzakis, Phys. Rev. X {\bf 4}, 021051 (2014).

\end{thebibliography}
\end{document}